\documentclass[12pt,a4paper]{article}
\usepackage{graphics}

\newcommand{\be}{\begin{equation}}
\newcommand{\ee}{\end{equation}}
\newcommand{\ba}{\begin{eqnarray}}
\newcommand{\ea}{\end{eqnarray}}
\newcommand{\dle}[1]{\label{#1}}

\newcommand{\dr}[1]{\ref{#1}}
\newcommand{\dc}[1]{\cite{#1}}
\newcommand{\paa}{\partial}

\newcommand{\gsim}{\raise.3ex\hbox{$>$\kern-.75em\lower1ex\hbox{$\sim$}}}
\newcommand{\lsim}{\raise.3ex\hbox{$<$\kern-.75em\lower1ex\hbox{$\sim$}}}
\def\IN{\relax{\rm I\kern-.18em N}}
\def\IR{\relax{\rm I\kern-.18em R}}

\font\cmss=cmss12 \font\cmsss=cmss12 at 7pt
\def\IZ{\relax\ifmmode\mathchoice
{\hbox{\cmss Z\kern-.4em Z}}{\hbox{\cmss Z\kern-.4em Z}}
{\lower.9pt\hbox{\cmsss Z\kern-.4em Z}}
{\lower1.2pt\hbox{\cmsss Z\kern-.4em Z}}\else{\cmss Z\kern-.4em Z}\fi}

\def\inbar{\,\vrule height1.5ex width.4pt depth0pt}
\def\IC{\relax\hbox{$\inbar\kern-.3em{\rm C}$}}

\newcommand{\pabar}{\not{\!{\partial}}}

\newcommand{\curr}{j_\mu (x)}

\newcommand{\Journal}[4]{#1 {\bf #2} (#4) #3}
\newcommand{\PRD}{Phys.Rev.D}
\newcommand{\PLB}{Phys.Lett.B}
\newcommand{\NPB}{Nucl.Phys.B}

\begin{document}

\title{Exploiting duality in a toy model of QCD at \\
non-zero temperature and chemical potential:
the massive Thirring
model, sine-Gordon model and Coulomb gases. }

\author{D.A. Steer
\\
D\'epartement de Physique Th\'eorique, 24 Quai Ernest Ansermet,
\\Universit\'e de Gen\`eve, 1211 Gen\`eve 4, Switzerland
\\ E-mail: {\tt Daniele.Steer@physics.unige.ch}
\\[0.5cm]
A. G\'{o}mez Nicola,
\\
Departamento de F\'{\i}sica Te\'orica, Universidad Complutense,
\\
28040, Madrid, Spain
\\ E-mail: {\tt gomez@eucmax.sim.ucm.es}
\\[0.5cm]
T.S. Evans and R.J. Rivers
\\
Blackett Lab., Imperial College, Prince Consort Rd.,
\\London, SW7 2BW, U.K.
\\  E-mails: {\tt t.evans@ic.ac.uk} and {\tt r.rivers@ic.ac.uk}
}


\maketitle

\abstract{ We focus on the massive Thirring model in 1+1
dimensions at finite temperature $T$ and non-zero chemical
potential $\mu$, and comment on some parallels between this model
and QCD.  In QCD, calculations of physical quantities such as
transport coefficients are extremely difficult.  In the massive
Thirring model, similar calculations are greatly simplified by
exploiting the duality which exists with the sine-Gordon model and
its relation, at high $T$, to the exactly solvable classical
Coulomb gas on the line.}

\section{Introduction and motivation}

The massive Thirring (MT) model in 2D Euclidean space with metric $(+,+)$
is our toy model of QCD.  It is described by the fermionic Lagrangian
\be
 {\cal{L}}_{MT} [\bar\psi,\psi] =  i\bar{\psi} (\pabar - m_0)\psi +
 \frac{1}{2}g^2  j_\mu (x) j^\mu (x) + \mu j_0(x),
\dle{Th}
 \ee
where
$\curr=\bar\psi (x) \gamma_{\mu} \psi (x)$ is the conserved current and
$\mu$ the chemical potential.  As in QCD,
(\dr{Th}) is only invariant under chiral
transformations
$\psi \rightarrow e^{i a \gamma_5} \psi $
if $m_0 = 0$, and
later we will study chiral symmetry restoration in this model
as $T\rightarrow\infty$, as a function of $\mu$ and coupling constant $g^2$.  We always take
$g^2 > 0$ so that
the
attractive interaction gives fermion anti-fermion bound states which correspond
to hadrons in QCD.

The sine-Gordon (SG) model,
on the other hand,
with Lagrangian
\be
{\cal L}_{SG}[\phi] = \frac{1}{2} \paa_\mu \phi  \paa^\mu \phi -
\frac{\alpha_0}{\lambda^2} \cos \lambda \phi ,
\dle{SG}
\ee
will play the part of
the effective chiral bosonic lagrangian
for low energy QCD.
There are an infinite number of degenerate vacua
$\phi_v = 2 n \pi/\lambda$ ($n \in \IZ$) and hence kink solutions.
These correspond to skyrmions (baryons) in low energy QCD.  Note that the
Lagrangian (\dr{SG}) is invariant under $\phi \rightarrow \phi +
\phi_v$ (the counterpart of isospin symmetry in QCD), whilst the potential
term breaks explicitly the symmetry $\phi \rightarrow \phi + a$,
which is the counterpart of the chiral symmetry in the MT
model.

At $T = \mu = 0$ the SG and MT models are
equivalent \dc{Coleman,Naon} and dual provided
\be
\frac{\lambda^2}{4\pi}= \frac{1}{1+g^2/\pi}.
\dle{eq}
\ee
Thus perturbative calculations in one theory tell us about
non-perturbative effects in the other.
Further, the weak identity
\be
\curr \leftrightarrow \frac{\lambda^2}
{2\pi} \epsilon_{\mu \nu}  \paa_\nu \phi(x)
\ee
shows that fermions in the MT model correspond to kinks in the SG model
(cf baryons and skyrmions in QCD).
 The bound
states in the MT model correspond to bosons and kink anti-kink breather
solutions in the SG model.

How are these links affected when $T>0$ and $\mu \neq 0$? One can
show\cite{Belg,GS} that at $T>0$, $\mu = 0$ the partition
functions of the two models are the same, $Z_{SG}(T,\mu = 0) =
Z_{MT}(T,\mu=0)$.  When $\mu \neq 0$, the equivalence $
Z_{SG}(T,\mu) = Z_{MT}(T,\mu)$ holds\dc{GS}, where $Z_{SG}(T,\mu)$
is now the partition function for the SG model with a topological
term which counts the number of kinks minus anti-kinks; ${\cal
L}_{SG}(\mu) = {\cal L}_{SG} - \mu \frac{\lambda}{2\pi} \frac{\paa
\phi}{\paa x}$.

\section{Massive Thirring model as a Coulomb Gas}

At any $T$ the MT model is not only equivalent to the SG model, but also to
non-relativistic particles of charge $\pm q$ (a neutral Coulomb gas (CG))
on a cylinder, circumference $T^{-1}$.
When the
cylinder collapses to a line  at temperatures $T \gsim m$, the
equivalent\dc{GRS} 1D neutral CG can be solved
exactly.

To see the equivalence, recall some of the properties of a 1D CG
at a temperature $T$.  The potential which binds the charges $q_i$
(=$\pm q$) at positions $x_i$ is $V(x_i,x_j) = -2 \pi q_i q_j |x_i - x_j|$, so
the grand canonical partition function for the system confined
on a line of length $L$ is
\be
\Omega (z,T,q,L)=\sum_{N=0}^{\infty}
\frac{ z^{2N}}{(N!)^2}
\left(\prod_{i=1}^{2N}\int_0^L dq_i\right)
\exp\left[2\pi q^2 T^{-1}\sum_{1\leq j<i\leq 2N}
\epsilon_i\epsilon_j \vert q_i-q_j\vert\right]
\label{coulparfun}.
\ee
Here $\epsilon_i=1$ for $i\leq N$ and $\epsilon_i=-1$ for $i>N$.
This classical problem can be solved
analytically\dc{lenard1,lenard2}. As a
result one finds that, at low $T$, the
CG behaves as a free gas of `molecules' made up of $+-$ charge pairs
bound together whereas,
at high $T$, the charges are deconfined and the pressure is large.

How is the MT model partition function related to (\dr{coulparfun})?
A hint comes from
noting that in perturbation theory about $\alpha_0 = 0$, $Z_{SG}(T)$
is a sum of terms each of which contains the expectation value of
products of $\cos \lambda \phi$'s.
Written in exponential form, these are just products of free massless boson
propagators which behave\dc{GRS} as $\Delta(T) \sim T |x|$ for
$T \gsim m$.
Thus one indeed gets a series of the form (\dr{coulparfun}).  More exactly,
\be
Z_{MT}(T,L) = Z_{0}^F(T,L) \Omega (z=f(m,T,g^2), T,q \propto T,L),
\dle{reln}
\ee
where $Z_{0}^F(T,L)$ is the partition function for free massless
fermions and $f$ is a specified function\dc{GRS} whose form is
unimportant here.  The important point to note from (\dr{reln}) is
that $q\propto T$. This implies that the CG behaviour above is
reversed. Furthermore one can show\dc{GRS} that the CG charges
correspond in the MT model to `chiral' charges $ \pm q
\leftrightarrow \hat{\sigma}_{\pm} = \hat{\bar{\psi}} (1 \pm
\gamma_5) \hat{\psi}$. Hence we are lead to an interesting picture
of the chiral symmetry properties of the MT model: at {\it low}
$T$ the system is in a `plasma phase' of free $\hat{\sigma}_{\pm}$
charges and  chiral symmetry is broken.  At {\it high} $T$ the
system is in the `molecular phase' in which the
$\hat{\sigma}_{\pm}$ bind into chirally invariant `molecules'
$\hat{\sigma}_+ \hat{\sigma}_-$ so that chiral symmetry is
restored.

This understanding can be verified by calculating the chiral condensate $\ll
\bar{\psi} \psi \gg$.
The result is plotted in Fig.\dr{con} for
different values of $g^2$.  Note that if $g^2=0$ chiral symmetry can
never be restored since there is not sufficient energy to bind the
molecules.  For larger $g^2$ chiral symmetry is restored asymptotically
as $T \rightarrow \infty$.
\begin{figure}[t]
{\centerline{\includegraphics{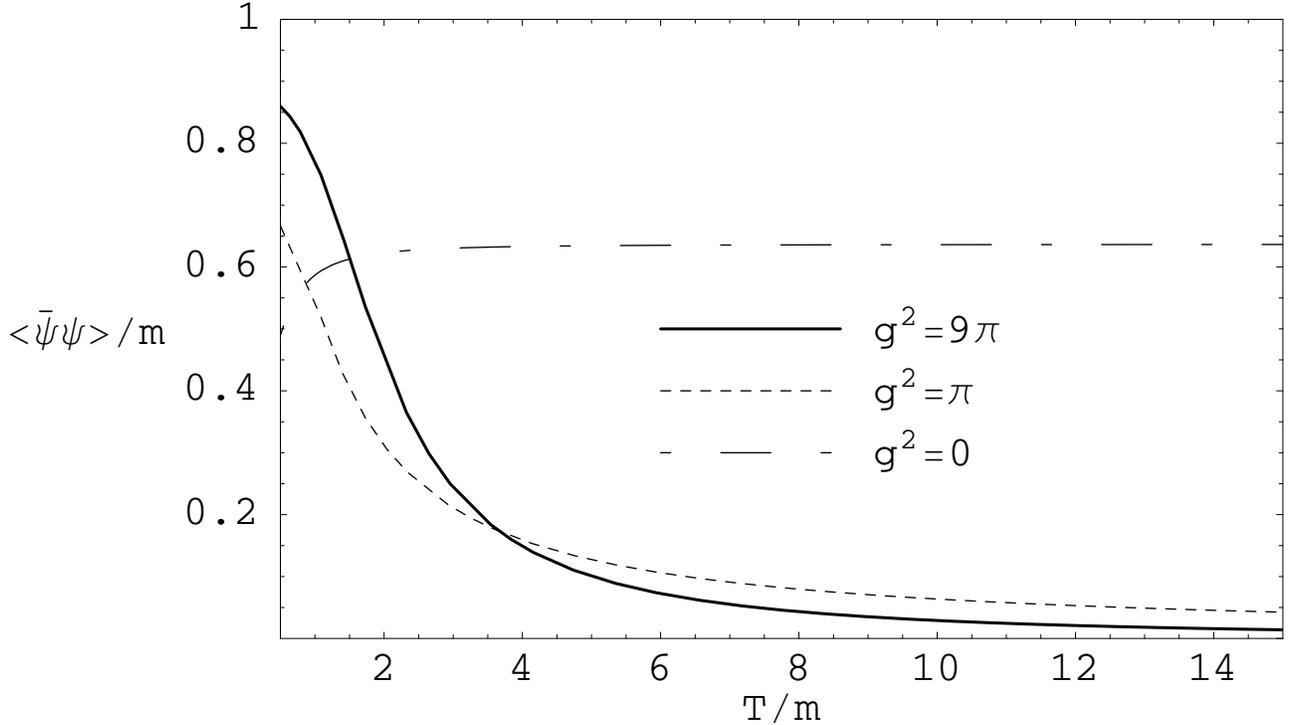}}} 
\caption{The Fermion condensate as a function of $T$ for different
$g^2$.} \label{con}
\end{figure}
Similarly one can calculate the pressure exactly in this $T\gsim m$
limit: it is given by $P_{MT} = \pi T^2/6 + P_{CG}(g^2,T)$ where the
first term is the contribution from the free massless fermions in
(\dr{reln}), and the CG part is given by
\be
P_{CG} = \frac{2\pi T^2}{1+g^2/\pi} \gamma_0 \left[ \frac{m^2}{4 \pi T^2}
\left( 1 + \frac{g^2}{\pi} \right) \left( \frac{2T}{m}
\right)^{\frac{1}{1+ g^2/\pi}} \right],
\ee
where $\gamma_0$ is the highest eigenvalue of the Mathieu differential
equation\dc{GRS}.  We stress that this result is non-perturbative and
exact for $T \gsim m$ --- it would be interesting to compare the
result with an order by order calculation in perturbation theory.
Similar comments apply for $\mu \neq 0$ when the CG picks up a
contribution from an (imaginary) external electric field\dc{GRS}; one
can then, for example, obtain the net averaged fermion density $\rho(T,\mu)$
exactly.

\section{Transport coefficients in the MT model}

Finally one can try to calculate transport coefficients in the MT
model, exploiting as much as possible the duality with the SG model
and the link with a 1D CG.  As an example of the power of this
approach, consider the response of the MT model to an external classical
electro-magnetic potential $A^{\mu}_{Cl}$.  From linear response theory
\be
\ll \delta j_{\mu}(x,t) \gg = -i \int_{-\infty}^{\infty} dt'
\int_{-\infty}^{\infty} dx' A_0^{cl}(x',t') \ll [j_{\mu}(x,t) , j_0(x',t') ] \gg \theta(t-t'),
\label{cond}
\ee
which, on using the duality with the SG model gives the conductivity (in
momentum space)\dc{EGRS}
\be
\delta j_1(k_0,k) = \sigma(k_0,k) E^{Cl}(k_0,k) = \frac{1}{1+g^2/\pi} k_0
D_R^{SG}(k_0,k) E^{Cl}(k_0,k)
\ee
Here $D_R^{SG}(k_0,k)$ is the retarded {\it bosonic} propagator of
the SG model: that is, we have rewritten the fermionic
current-current correlator in terms of bosonic collective modes (a
quasi-particle propagator).  The existence of such collective
modes would not be obvious without duality, but is confirmed by
expanding the r.h.s. of (\ref{cond}) in powers of the renormalised
$\alpha$ (fermionic mass). The dominant free bosonic term is the
conductivity of the {\it massless} Thirring model (i.e.\ a quantum
wire). For a finite length wire with appropriate boundary
conditions the IR singularities of this leading term (now exact)
are controllable and calculable\cite{stone}, and the predicted
conductance can be observed experimentally. For our {\it massive}
Thirring model, analysis\dc{EGRS} of $D_R^{SG}(k_0,k)$ shows that
it, in addition, contains the full bosonic self-energy which,
through duality, is related to the fermion condensate $\ll
\bar{\psi} \psi \gg$ and the net average fermion density
$\rho(T,\mu)$, which we have already calculated. Further results
will be presented elsewhere\dc{EGRS}.

\section{Conclusions}

We have tried to summarise some intriguing relations which hold
between the $T>0$, $\mu \neq 0$ MT model, and the SG model and Coulomb
gases.  The particular link with a 1D CG for $T \gsim m$ has enabled many
thermodynamic quantities to be obtained analytically as discussed in
section 2, and we also made some first steps at tackling transport
coefficients in section 3 in terms of the dual modes.

The relationship with the CG also provided an interesting
interpretation of chiral symmetry restoration in the MT model in terms
of binding of `chiral charges' $\hat{\sigma}_{\pm}$.  Can any of these
pictures be of relevance to QCD, for which the analogue Coulomb gas is one of
 monopoles\dc{Jai}?  We believe that a tentative answer to that question is
yes, particularly for the understanding of chiral symmetry restoration in QCD.

\section*{Acknowledgements}

D.A.S., T.S.E and R.J.R would like to thank PPARC  for financial support.
A.G.N would like to thank CICYT.
This work was also supported, in part, by the ESF.

\end{document}